 \documentclass[12pt,preprint]{aastex}

\begin{document}

\title{Terrestrial Ozone Depletion Due to a Milky Way Gamma-Ray Burst}


\author{Brian C. Thomas\altaffilmark{1}, Charles H. Jackman\altaffilmark{2}, 
	Adrian L. Melott\altaffilmark{1}, Claude M. Laird\altaffilmark{1,4},
	Richard S. Stolarski\altaffilmark{2}, 
	Neil Gehrels\altaffilmark{3}, John K. Cannizzo\altaffilmark{3}, and
	Daniel P. Hogan\altaffilmark{1}
	}		

\altaffiltext{1}{University of Kansas, Department of Physics and Astronomy \\
		 1251 Wescoe Hall Dr, Room 1082 \\ 
        	 Lawrence, KS  66045-7582; bthomas@ku.edu, melott@ku.edu, claird@ku.edu }
\altaffiltext{2}{Laboratory for Atmospheres \\
		 NASA Goddard Space Flight Center \\
		 Code 613.3 \\	
		 Greenbelt, MD 20771; jackman@assess.gsfc.nasa.gov, stolar@polska.gsfc.nasa.gov}
\altaffiltext{3}{Laboratory for Astroparticle Physics \\
	  	 NASA Goddard Space Flight Center \\
		 Code 661; gehrels@milkyway.gsfc.nasa.gov, cannizzo@milkyway.gsfc.nasa.gov \\	
		 Greenbelt, MD 20771; gehrels@milkyway.gsfc.nasa.gov, cannizzo@milkyway.gsfc.nasa.gov}
\altaffiltext{4}{Also at Haskell Indian Nations University}

%
%
%
%
%
%

\begin{abstract} 
Based on cosmological rates, it is probable that at least once in the last Gy the Earth has been irradiated by a gamma-ray burst in our Galaxy from within 2 kpc.  Using a two-dimensional atmospheric model we have performed the first computation of the effects upon the Earth's atmosphere of one such impulsive event.  A ten second burst delivering $100~ \mathrm{kJ/m}^2$ to the Earth penetrates to the stratosphere and results in globally averaged ozone depletion of 35\%, with depletion reaching 55\% at some latitudes.  Significant global depletion persists for over 5 years after the burst.
This depletion would have dramatic implications for life since a 50\% decrease in ozone column density results in approximately three times the normal UVB flux.
Widespread extinctions are likely, based on extrapolation from UVB sensitivity of modern organisms.  Additional effects include a shot of nitrate fertilizer and  $\mathrm{NO_2}$ opacity in the visible providing a cooling perturbation to the climate over a similar timescale.  These results lend support to the hypothesis that a GRB may have initiated the late Ordovician mass extinction \citep{mel04}.
\end{abstract}


\section{Introduction}
Gamma-ray bursts (GRB) within our galaxy have been suggested as a possible threat to life on Earth \citep{th95,sw02,dar02,mel04}.  Some effects similar to that due to a nearby supernova (SN) are expected, especially depletion of stratospheric ozone due to ionization caused by incident gamma radiation.  GRB are rarer than supernovae, but their greater energy output results in a larger region of influence and hence they may pose a greater threat.  It is likely \citep{mel04,der05} that in the last Gy a GRB has occurred close enough to have dramatic effects on stratospheric ozone, leading to detrimental effects on life through increases in solar ultraviolet (UV) radiation which is strongly absorbed by ozone.  A major issue is the timescale for atmospheric chemistry: most of the GRB fluence comes in seconds or minutes versus months for supernovae.

In order to gain more detailed and accurate insight into these expected effects we have performed computations using the Goddard Space Flight Center (GSFC) two-dimensional atmospheric model.  This model has been used previously to investigate the atmospheric effects of SN \citep{geh03}.  While we build on that work, the computations discussed here are significantly more challenging due to the extremely short duration and greater energy output of GRB in comparison to SN.

\section{Methods}
\label{sec-methods}
We take as ``typical'' a GRB with power $5\times10^{44}~\mathrm{W}$ (isotropic equivalent) and duration 10 s, whose gamma-ray spectrum is described by the Band spectrum \citep{band93}. These assumptions are drawn from observations and are not dependent upon beaming angle. 
We have determined the depletion of ozone for such a GRB beamed at the Earth from a distance of 2 kpc, delivering to the Earth a total fluence of $100~ \mathrm{kJ/m}^2$.  This distance corresponds to that of a probable nearest ``typical'' GRB in the last Gy, based on conservative assumptions  \citep{mel04}.

The prompt effect of this burst at the Earth's surface is a ``flash'' of UVB radiation with power $\sim 20~ \mathrm{W/m^2}$ \citep{smith04}.  
This is about seven times the intensity at the Earth's surface on a bright, sunny day, but is brief and so is not likely to have a major effect on life.  Longer term effects include ozone depletion and the resulting increase in solar UVB flux, which we begin to explore here.

We have not included the effects of any ultra high energy ($>10^{18}~\mathrm{eV}$) cosmic rays from a GRB \citep{der04,wax04a,wax04b,der05}, due to uncertainty as to whether and at what energies GRB may produce such particles.  Lower energy cosmic rays which may be produced are deflected at our assumed distance by the galactic magnetic field.  Inclusion of ultra high energy cosmic rays would measurably increase the ozone-destroying energy budget.  In this case, we would also expect production of some radioisotopes from cosmic ray spallation off of atmospheric constituents, as well as a terrestrial burst of muons.

\subsection{Atmospheric Model}
The GSFC 2D model is described in \citet{djs89}; \citet{jack90}; and \citet{cdj94}. The model's two dimensions are latitude and altitude (ranging up to about 116 km).  The latitude range is divided into 18 equal bands and extends from pole to pole.  The altitude range includes 58 evenly spaced logarithmic pressure levels (approximately 2 km spacing).   A lookup table is used for computation of photolytic source term, used in calculations of photodissociation rates of atmospheric constituents by sunlight \citep{jack96}.  Winds and small scale mixing are included as described in \citet{flem99}.  For this study, we have removed anthropogenic compounds such as CFCs.

We have employed two versions of the atmospheric model.  One is intended for long term runs (many years) and includes all transport mechanisms (e.g., winds and diffusion); it has a time step of one day and computes daily averaged constituent values.  The second is used for short term runs (a few days) and calculates constituent values throughout the day and night, but does not include transport.  Previously, this version has been used with a time step of 225 seconds \citep{jack01}.  In the current study we have used a time step of one second in order to allow for spreading our GRB gamma radiation input over several time steps.

Gamma-rays are introduced in the model in a manner similar to that described in \citet{geh03}.  In that study, gamma-rays were included using the spectrum of SN 1987A.  In the present study, the gamma-ray differential photon count spectrum used is that of \citet{band93}, which consists of two smoothly connected power laws.  We use the following typical values for the break energy and power law indices, respectively: $E_0 = 187.5~ \mathrm{keV}$, $\alpha = -0.8$, $\beta = -2.3$ \citep{preece00}. The total incident energy is scaled to our desired value (in this study, corresponding to a fluence of $100~ \mathrm{kJ/m}^2$)
The total photon flux in each of 66 evenly spaced logarithmic energy bins, ranging $0.001 \leqq E \leqq 10 ~ \mathrm{MeV}$, is obtained by integrating the Band spectrum for each bin.

\subsection{Simulations}
\label{sec-sims}
All simulation runs used for analysis were begun with initial conditions obtained from a long-term (roughly 40 years) run intended to bring the model to equilibrium.  Constituent values from this run are read in by the 1s time step version of the model which runs for 7 days (beginning at noon), either with or without input of gamma-rays.  Runs including gamma radiation treat the burst as a step function at noon on day 4, with duration 10 s. In the current study, the burst is input in late March (near the spring equinox) with incident angle 0 degrees (equatorial).  (Forthcoming studies will investigate the effects of varying intensity, incidence angle, and times of year at which the burst occurs.)  Day 4 is the middle of the short term run, which allows for ``warm up'' in the model and insures that relevant chemistry is accurately computed over many time steps after the burst.  Constituent values from this type of run are then read in by the 1 day time step version which is run for 20 years in order to investigate long term effects and determine how long the atmosphere takes to return to equilibrium, pre-burst conditions.

Analysis is then performed by comparing such a combined base-short-long run without gamma-ray input to such a run with the burst included.  Ozone depletion and other effects are computed by comparing these two runs.

\section{Results}
Stratospheric ozone is lost through several catalytic reactions involving oxygen-, nitrogen-, hydrogen-, chlorine-, and bromine-containing gases.  The constituents in the stratosphere are generally grouped into ``families'' such as $\mathrm{O_x}$ ($\mathrm{O_3}$, O, $\mathrm{O(^1D)}$), $\mathrm{NO_y}$ (N, NO, $\mathrm{NO_2}$, $\mathrm{NO_3}$, $\mathrm{N_2O_5}$, $\mathrm{HNO_3}$, $\mathrm{HO_2NO_2}$, $\mathrm{ClONO_2}$, $\mathrm{BrONO_2}$), $\mathrm{HO_x}$ (H, OH, $\mathrm{HO_2}$), $\mathrm{Cl_y}$ (chlorine-containing inorganic molecules), and $\mathrm{Br_y}$ (bromine-containing inorganic molecules), which allow for efficient computation of the chemistry and transport effects.  We assume that there were no anthropogenic sources for any of these families, as human influence has been negligible for most of geologic time.

In the case of a large input of gamma rays to the atmosphere $\mathrm{NO_y}$ compounds (most importantly NO and $\mathrm{NO_2}$) are created through dissociation of $\mathrm{N_2}$ in the stratosphere which then reacts quickly with $\mathrm{O_2}$ to generate NO.  Subsequent reactions create $\mathrm{NO_2}$ and other compounds.  Together, these react catalytically to deplete $\mathrm{O_3}$  through the cycle
\begin{eqnarray} 
\mathrm{ NO + O_3 \rightarrow NO_2 + O_2 } \\
\mathrm{ NO_2 + O \rightarrow NO + O_2 } \\
\mathrm{ net: O_3 + O \rightarrow O_2 + O_2 }
\end{eqnarray}

Note that NO is not consumed in this cycle and the net result is the destruction of $\mathrm{O_3}$ and production of $\mathrm{O_2}$.  Other reactions can complicate this cycle, such as destruction of $\mathrm{NO}$ by reaction with $\mathrm{N}$; production of $\mathrm{O_3}$ through reactions of $\mathrm{NO}$ with $\mathrm{HO_2}$; and interference of $\mathrm{NO_y}$ with other families (chlorine-, bromine-, and hydrogen-containing constituents) which reduces the ozone depletion from these families.  Some uncertainties in the atmospheric model's treatment of this cycle are discussed in Section~\ref{sec-uncert}.
 
The primary results of our simulations are increases in $\mathrm{NO_y}$ and decreases in $\mathrm{O_3}$.  Ozone column densities can then be used to calculate the resulting UVB flux at the Earth's surface and from this information biological effects can be estimated.  UVB is particularly dangerous to organisms because DNA is damaged by absorption in this wavelength range.

\begin{figure}
\plotone{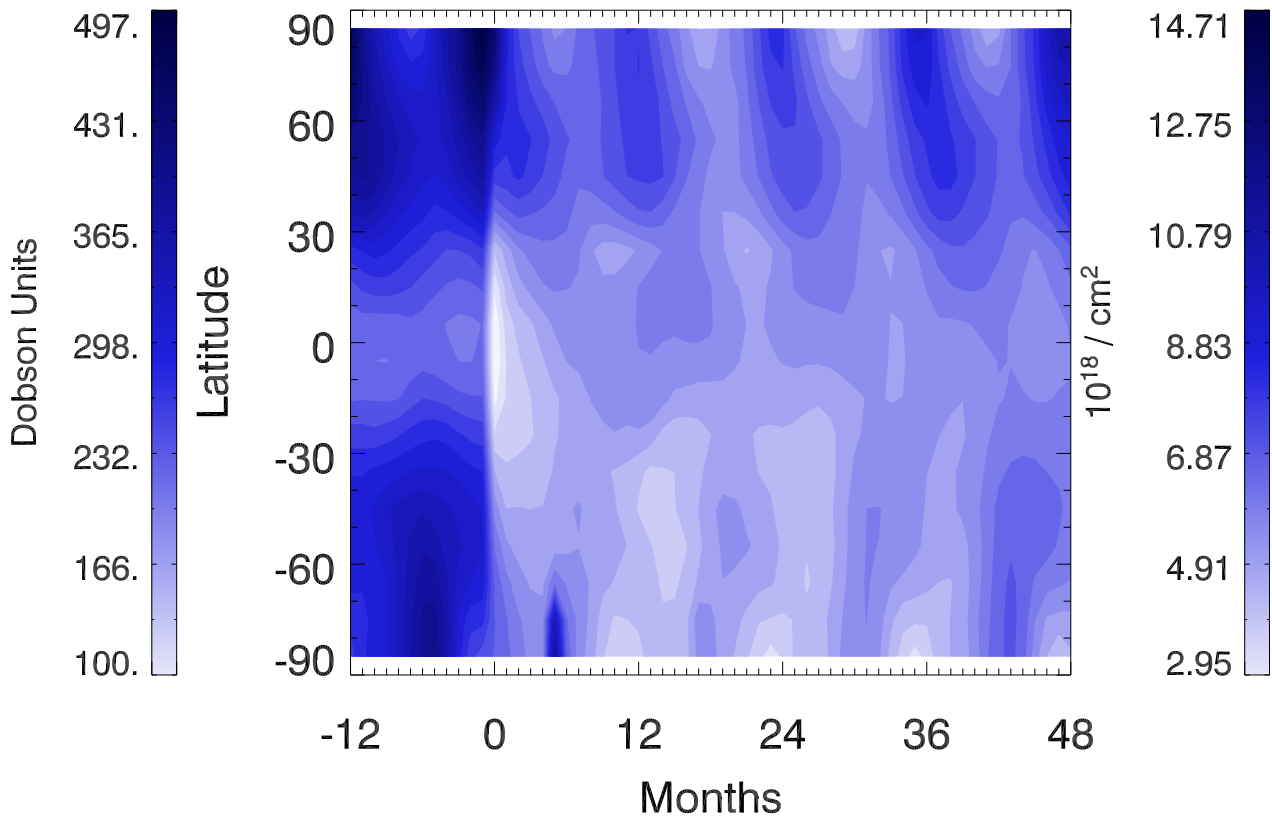}
\caption{Column density of $\mathrm{O_3}$ with scales for both Dobson units (left) and $10^{18}~\mathrm{cm^{-2}}$ (right). (Plotted from one year before the burst to four years after.) \label{fig:O3_coldens}}
\end{figure}

\begin{figure}
\plotone{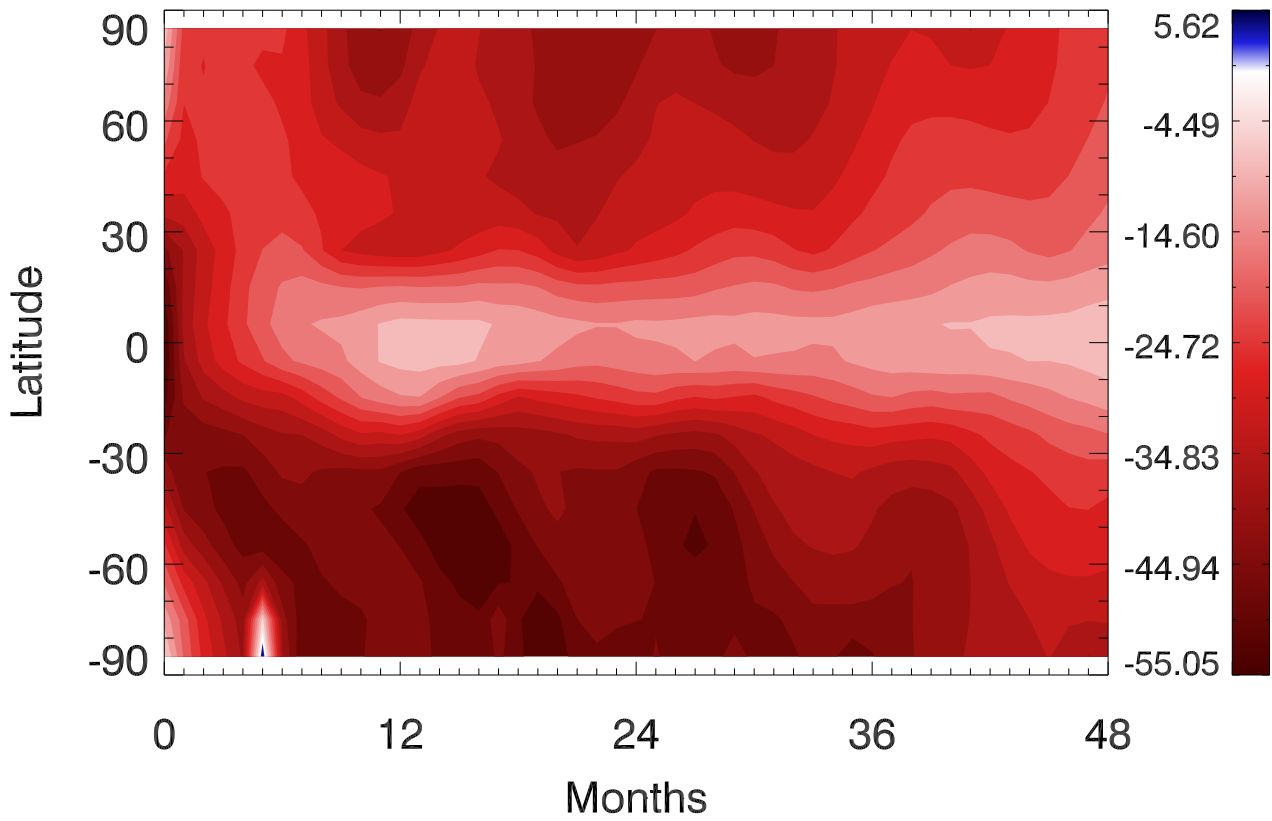}
\caption{Pointwise percent change in column density of ozone (comparing runs with and without burst). \label{fig:O3_perchng}}
\end{figure}

Results of our modeling are shown in Figures~\ref{fig:NOy_coldens},~\ref{fig:O3_coldens} and~\ref{fig:O3_perchng}.
We have modeled the effects of $100~ \mathrm{kJ/m}^2$ total incident gamma-ray fluence, input as described in Sect.~\ref{sec-sims}.  This corresponds to our ``typical'' GRB located at about 2 kpc.
Figure~\ref{fig:NOy_coldens} shows the vertical column density of $\mathrm{NO_y}$ at each latitude over time.  The burst is input at time 0.
Figure~\ref{fig:O3_coldens} shows the vertical column density of $\mathrm{O_3}$.  Included are scales in both Dobson units (the usual unit of ozone column density) and $10^{18}~\mathrm{cm^{-2}}$.  A Dobson unit describes the thickness of a column of ozone at STP and is defined as $1~\mathrm{DU} = 0.01~\mathrm{mm}$ thickness (or, $1~\mathrm{DU} = 2.69\times10^{18}~\mathrm{cm^{-2}}$).
Figure~\ref{fig:O3_perchng} shows the percent difference at a given location (between a run with gamma-ray input and one without) in vertical column density of $\mathrm{O_3}$. Immediate depletion of ozone is evident.  
Due to their qualitative similarity, we have chosen not to plot here changes in $\mathrm{NO_y}$.  Maximum increase in $\mathrm{NO_y}$ is largely coincident with maximum decrease in $\mathrm{O_3}$.  The maximum increase in $\mathrm{NO_y}$ at a given location is approximately 30-fold for this case.


Several features in these plots are worth noting.  First, as is seen in Fig.~\ref{fig:O3_coldens} and~\ref{fig:O3_perchng}, depletion of ozone is initially greatest at the equator (where the incident flux is highest), becoming greatest toward the poles within a year or so.  
Larger ozone depletions at the poles are primarily due to the long lifetime of the enhanced $\mathrm{NO_y}$ in the polar stratosphere. Figure~\ref{fig:O3_perchng} gives a somewhat exaggerated impression of the effect of depletion at the poles, since ozone is initially high there.  The enhanced $\mathrm{NO_y}$, including $\mathrm{HNO_3}$, will lead to an enhancement of nitric acid trihydrate (NAT) polar stratospheric clouds (PSCs).  These NAT PSCs facilitate heterogeneous reactions that result in greater ozone depletion by halogen (chlorine and bromine) constituents.  This is especially true in the south polar region where a stronger polar vortex with colder stratospheric temperatures is in place during the winter. This contributes to an asymmetry which would be peculiar to the present-day configuration of continents.  A much larger effect which contributes to the polar asymmetry is the time of year at which the burst occurs, since ozone concentrations at the poles exhibit large seasonal variations. 
 This asymmetry is largely north-south reversed for a burst in September rather than March (Thomas et al. 2005, in preparation).  We therefore conclude that present peculiarities of continent distribution (including the south polar vortex) are not a major source of uncertainty in this work.

Around 5-6 months after the burst there is a short-lived \emph{production} of ozone toward the south pole.  
Production occurs at the end of south polar night when a lack of photolysis has caused accumulation of $\mathrm{NO_y}$ constituents which are suddenly photolyzed as the sun rises, producing O which may then react with $\mathrm{O_2}$ to form $\mathrm{O_3}$.  

Globally averaged ozone depletion reaches about 35\% (at the start of the long-term run) and a maximum depletion of about 55\% occurs first at the equator immediately after the burst, and then again about 15 months after the burst, in the southern hemisphere.  Significant global depletion (10\% or more) lasts for over 5 years after the burst.

\begin{figure}
\plotone{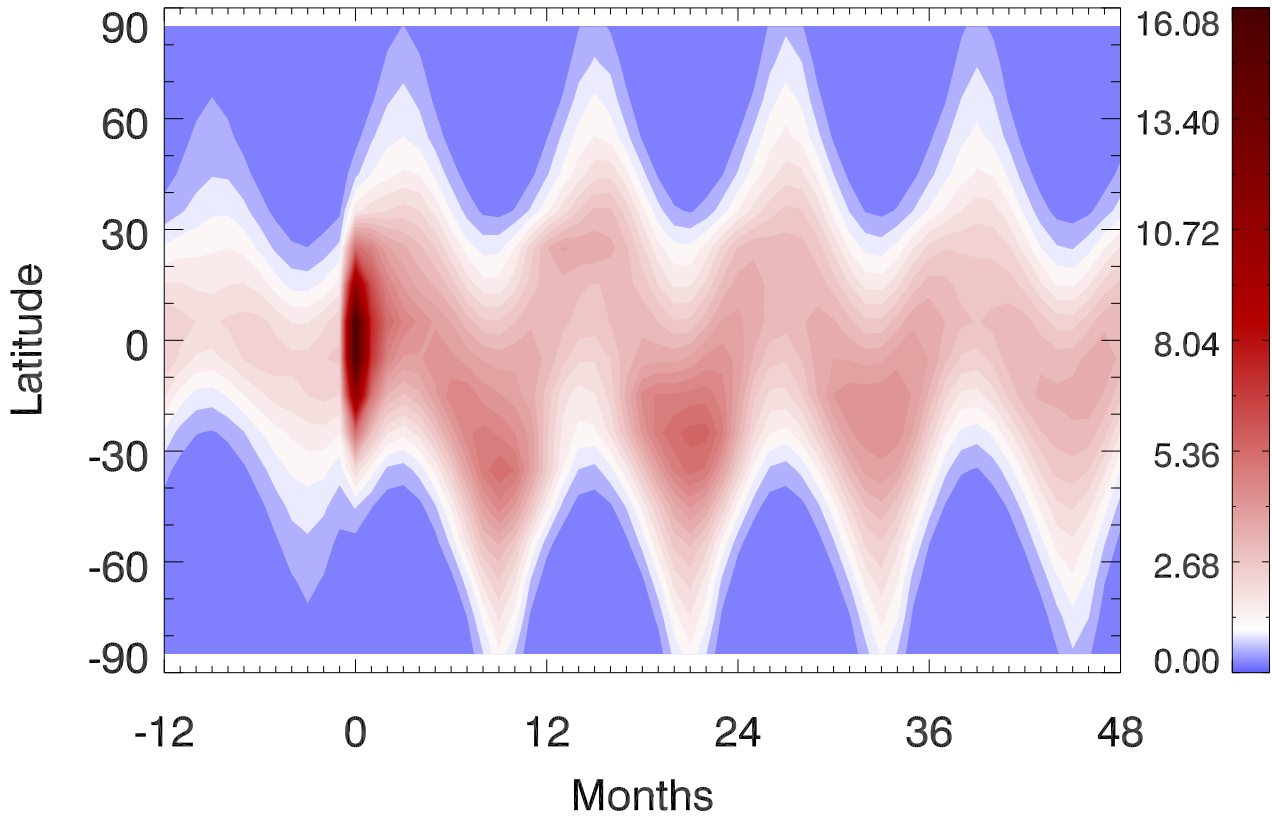}
\caption{Relative DNA damage (dimensionless), normalized by the annual global average damage in the absence of a GRB. (Plotted from one year before the burst to four years after.)  Note that white is set to 1.0.  This figure is available as an animation online at http://kusmos.phsx.ukans.edu/$\sim$melott/DNA\_damage\_MPEG.mpg . \label{fig:lat_DNA}}
\end{figure}

Figure~\ref{fig:lat_DNA} (available as an animation online at http://kusmos.phsx.ukans.edu/$\sim$melott /DNA\_damage\_MPEG.mpg) shows DNA damage estimated by convolving the daily average UVB flux at the ground with a biological weighting function \citep{set74,smith80}.  We include only ozone absorption effects on the UVB flux since the effect of scattering at these wavelengths is comparatively small.  We have normalized the plot by dividing the damage by the annual global average damage in the absence of a GRB. Greater DNA damage probability is evident at low latitudes. This is due to combination of the $\mathrm{O_3}$ depletion effects with the sun incidence angle, length of day, etc.  
We have performed other runs with different GRB incidence latitudes, times of year, etc., to be discussed elsewhere (Thomas et al. 2005, in preparation) and found that the concentration of computed DNA damage to low-mid latitudes is a general feature of the GRB hypothesis. 
One might think that this damage would be countered by a greater evolved UVB-resistance in organisms at low latitudes.  However, at least for modern organisms, there is no evidence that temperate zone phytoplankton are any more UVB-resistant than Antarctic plankton \citep{prez04}.  Thus, one might predict that greater ecological damage and extinction would be likely near the equator.  It is interesting that the late Ordovician mass extinction seems to be alone in having recovering fauna preferentially derived from high-latitude survivors \citep{jab04,shee01}.
 
\section{Uncertainties}
\label{sec-uncert}
Our ionization profiles are computed using simple energy-dependent attenuation coefficients, instead of a full radiative transfer calculation.  This technique is implemented following \citet{geh03} with the primary modification being the functional form of the spectrum.  That study found good agreement between their calculations and a full radiative transfer model.  We find that ionization due to the gamma-ray input peaks around $30~\mathrm{km}$ elevation, which is in agreement with \citet{geh03} and \citet{smith04}.

Effects of prompt redistributed UV are not included in the model.  This will both produce and destroy ozone.  We have performed a simple test of the magnitude of such effects by increasing the solar flux by $10^4$ times for 10s to simulate the redistributed UV.  The resulting ozone depletion is a few percent greater.  Also, in some cases the ionizing fluence from the GRB afterglow may be as large as that from the burst.  So, at worst our results are somewhat conservative.  As discussed in \citet{mel04} and \citet{der05}, there is enough uncertainty in the GRB rate at low redshift that the fluence from the probable nearest burst could be one order of magnitude greater than we model here.

A discussion of some uncertainties in the production of $\mathrm{NO_y}$ compounds is presented in \citet{mel04}.  In particular, reactions involving exited state nitrogen atoms, $\mathrm{N(^2D)}$, are not included in the atmospheric model.  Comparison with simplified, off-line computations using various ratios of $\mathrm{N(^2D)}/\mathrm{N(^4S)}$ indicate that for our present case the effect on NO production of including excited state N atoms is small.  Though results depend on the assumed temperature at which the reactions occur, for a reasonable range of temperatures (230-270 K), there is little variation of NO production with increased concentrations of excited state N atoms, and there is generally less than a factor of two difference between the off-line computations and model results.  An additional complication is the effect of reactions involving HNO which may limit the production of $\mathrm{NO_y}$.  Such reactions are not included in the model, but we estimate that this is a small effect.

The GSFC two-dimensional atmospheric model is empirically based and its dynamics are not coupled to the significantly changing constituent levels and accompanying heating.  This fact introduces some uncertainty in the transport of constituents.  As mentioned before, variations due to input latitude and time of year are likely more significant.

\section{Discussion}
A significant result of our modeling is that even for a short duration input of radiation, atmospheric effects are large and long-lived.  Expectations based on supernova studies indicated that a short duration input might not have such large effects \citep{geh03}.
Appearance of features such as the localized production of $\mathrm{O_3}$ highlights the need for detailed modeling of the effects of a GRB on the Earth's atmosphere, as complexities and feedbacks in the chemical processes can be important and simple scaling may not prove reliable.

\citet{mel04} summarizes studies of UVB sensitivity of various organisms (see also \citet{coc99}).  About 90\% of UVB is presently absorbed by atmospheric ozone.  Due to the sensitivity of DNA to this radiation, increases of only 10-30\% can have lethal effects on many organisms, especially phytoplankton, the base of the food chain.  Ozone depletions in the range of 50\%, as seen here, lead to roughly three times more UVB at the surface, which is clearly a possible candidate for causing mass extinctions.  Of course, we expect additional events from smaller burst fluences over the last Gy, less intense but still significant for the biosphere.

There are other effects.  The event described here could potentially produce of order $0.5~ \mathrm{g/m^2}$ mean global deposition of nitrates.  Biota are generally nitrate-starved, and this deposition may have eased the transition to land, which accelerated after the Ordovician.  This nitrate deposition may provide a geochemical signature which could serve as a test of our hypothesis, though this would be difficult due to the extreme water solubility of nitrates.  On the other hand, our hypothesis is falsifiable on geochemical grounds. That is, a layer of iridium (associated with impact events) or radioisotopes such as $^{244}\mathrm{Pu}$ (associated with SN events \citep{ell96}) would not be associated with our scenario.  Different radioisotopes could be generated by spallation if significant levels of cosmic rays are received (see Sect.~\ref{sec-methods}).  However, few would survive to the present for the late Ordovician mass extinction, which occurred 443 My ago. 

\begin{figure}
\plotone{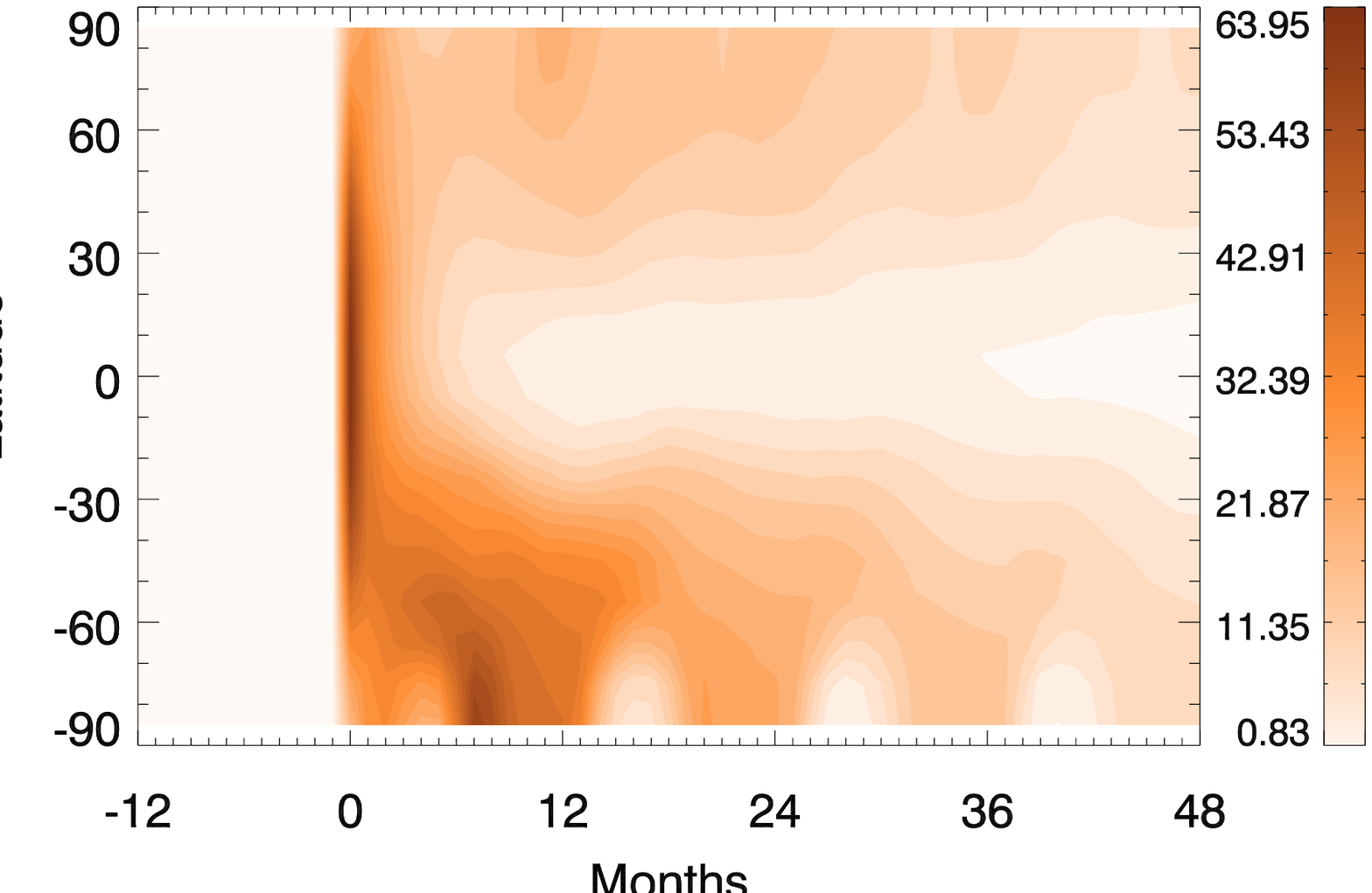}
\caption{Column density of $\mathrm{NO_y}$ in units of $10^{18}~\mathrm{cm^{-2}}$. (Plotted from one year before the burst to four years after.) \label{fig:NOy_coldens}}
\end{figure}

The Ordovician extinction is associated with a brief glaciation in the middle of a period of stable warm climate.  We speculate that there may have been a significant perturbation by the opacity of $\mathrm{NO_2}$, which would cut off a few percent (ranging up to 35\% for a month or so during polar fall) of solar radiation \citep{reid78}.  This would occur primarily at high latitudes, as can be seen in Fig.~\ref{fig:NOy_coldens}.  The removal of $\mathrm{O_3}$ (a greenhouse gas) also may cause some cooling, but this effect should be negligible compared to that due to the increase in $\mathrm{NO_2}$   We will provide more detail on these ideas in the near future.

\acknowledgments
A.L.M. wishes to thank J. Scalo, D. Smith, and J.C. Wheeler for useful conversations.
B.C.T. acknowledges support from Neil Gehrels at NASA Goddard Space Flight Center. 
B.C.T., A.L.M. and C.M.L. acknowledge support from NASA Astrobiology grant NNG04GM41G.
B.C.T. and A.L.M. acknowledge supercomputing support from the National Center for Supercomputing Applications.
D.P.H. acknowledges support from an Undergraduate Research Award through the Honors Program at the University of Kansas.

\clearpage

\clearpage

\clearpage

\clearpage

\end{document}